\documentclass[sigconf]{acmart}
\AtBeginDocument{%
  \providecommand\BibTeX{{%
    \normalfont B\kern-0.5em{\scshape i\kern-0.25em b}\kern-0.8em\TeX}}}
\usepackage{subfig}
\setcopyright{acmcopyright}
\copyrightyear{2018}
\acmYear{2018}
\acmDOI{10.1145/1122445.1122456}

\acmConference[FDG'20]{FDG'20: Foundations of Digital Games}{September 15--18, 2020}{Malta}
\acmBooktitle{FDG'20: Foundations of Digital Games,
 September 15--18, 2020, Malta}
\acmPrice{15.00}
\acmISBN{978-1-4503-XXXX-X/18/06}

\begin{document}

\title[Moment-to-moment Engagement Prediction through the Eyes of the Observer]{Moment-to-moment Engagement Prediction through the Eyes of the Observer: PUBG Streaming on Twitch}

\author{David Melhart}
\authornote{Both authors contributed equally to this research.}
\email{david@modl.ai}

\author{Daniele Gravina}
\authornotemark[1]
\email{daniele@modl.ai}

\author{Georgios N. Yannakakis}
\email{georgios@modl.ai}
\affiliation{\institution{modl.ai}}

\renewcommand{\shortauthors}{Melhart, Gravina and Yannakakis}

\begin{abstract}
Is it possible to predict moment-to-moment gameplay engagement based solely on game telemetry? Can we reveal engaging moments of gameplay by observing the way the viewers of the game behave? To address these questions in this paper, we reframe the way gameplay engagement is defined and we view it, instead, through the eyes of a game's live audience. We build prediction models for viewers' engagement based on data collected from the popular battle royale game \emph{PlayerUnknown's Battlegrounds} as obtained from the \emph{Twitch} streaming service. In particular, we collect viewers' chat logs and in-game telemetry data from several hundred matches of five popular streamers (containing over $100,000$ game events) and machine learn the mapping between gameplay and viewer chat frequency during play, using small neural network architectures. Our key findings showcase that engagement models trained solely on 40 gameplay features can reach accuracies of up to 80\% on average and 84\% at best. Our models are scalable and generalisable as they perform equally well within- and across-streamers, as well as across streamer play styles. 
\end{abstract}

\ccsdesc[500]{Applied computing~Computer games}
\ccsdesc[300]{Human-centered computing~User models}

\keywords{Machine learning, artificial neural networks, engagement, viewer modelling, streaming, battle royale games, PUBG}

\maketitle

\section{Introduction}

\begin{figure}[!tb]
\centering
\includegraphics[width=1.0\linewidth]{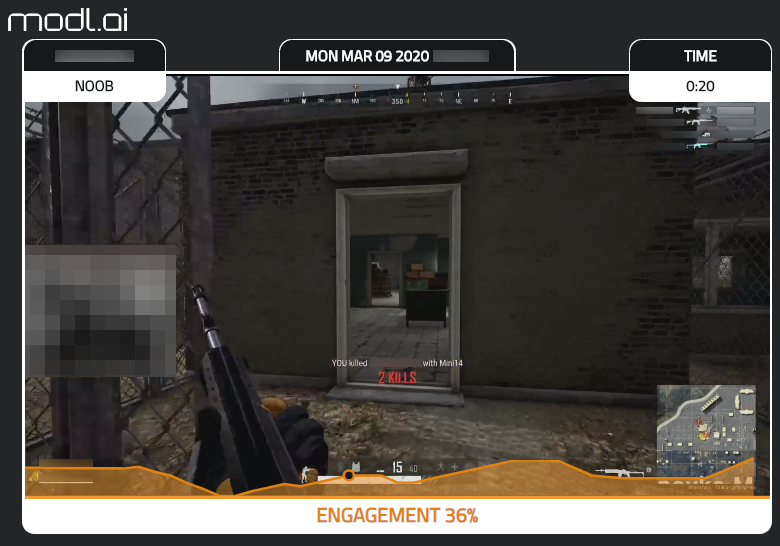}
\caption{Moment-to-moment engagement prediction of a live PUBG streamer. In the middle of the screen a replay of the game can be seen. The streamer's face is on the left (blurred out to preserve anonymity). Engagement is illustrated as a continuous orange line at the bottom of the video; its current value is displayed underneath. The streamer's name (also blurred in this example) and play style (i.e., \emph{Noob} in this example) are shown at the top left of the dashboard. See Section \ref{sec:clustering} for more details regarding the methods used to identify PUBG play styles. The video on the screenshot is obtained through the Twitch Developer API (fair use), PUBG is a registered trademark of PUBG CORPORATION.}\label{fig:modl} 
\end{figure}

The reliable estimation of the moment-to-moment gameplay engagement is arguably of utmost value for game development. Accurate proxies of engagement may not only enhance a game's monetisation strategy, they can also be used for rapidly testing games through artificial game-playing agents that are equipped with such engagement estimates. Artificial intelligence algorithms that are driven by reliable engagement models can, in turn, improve aspects of player experience and lead to the design of entirely new and engaging gameplay via game content generation \cite{yannakakis2018artificial}. 

With the advent of streaming services and the growing popularity of electronic sport competitions, the engagement of game spectators became increasingly important; if not as important as the engagement of players themselves. This is evidenced by the exponential growth of game streaming services such as \emph{Twitch}\footnote{\url{https://www.twitch.tv}} and \emph{Mixer}\footnote{\url{https://mixer.com/}} and the audience of games like \emph{PlayerUnknown's Battlegrounds} (PUBG Corporation, 2017) or \emph{Fortnite} (Epic Games, 2017) in recent years. 
Unconventionally in this study, instead of looking at the player's behaviour as a predictor of engagement we \emph{reframe} the modelling problem and look at gameplay engagement from the viewers' perspective. We define \emph{gameplay} as the state of a game that is experienced and \emph{engagement} as the active participation of viewers of gameplay. We thus assume there is an unknown mapping between gameplay---that is live streamed to viewers---and the engagement of the audience of that game.  

To test this hypothesis we utilise the popular video live streaming service \emph{Twitch} and obtain data from 5 popular streamers of the game \emph{PlayerUnknown's Battlegrounds---PUBG} (PUBG Corporation, 2017) as streamed on \emph{Twitch}. Importantly, three of those streamers are currently ranked within the top 10 PUBG streamer list---rank 1, 2 and 7---in terms of viewership. To construct models of moment-to-moment gameplay engagement in this initial study we investigate the relationship between critical events of the game and the corresponding frequency of messages in the chat feed. In particular, we use artificial neural networks that are able to predict gameplay engagement (as attributed to the viewers' chat frequency) at each critical event in the game (e.g., player death, head-shot, kill etc.). The derived models reach accuracies of up to 80\% on average and 84\% at best suggesting that gameplay events can form accurate predictors of viewer engagement and that viewer behaviour (through the frequency of chatting) can be attributed to gameplay. Our models are able to predict engagement within and across the five different streamers with similarly high accuracies showcasing the scalability and generalisability of the approach. Moreover, the models can accurately predict engagement---with accuracies up to 75-80\% on average---across three different PUBG play styles (\emph{Noob}, \emph{Explorer} and \emph{Pro}) which are identified through data clustering methods. The outcome of this work is a continuous prediction of engagement (engagement line) and play style for any given live PUBG video that is streamed (see Figure \ref{fig:modl}).

This paper is novel in several ways. First it approaches gameplay engagement from a third-person (viewer) rather than a first-person (player) perspective, as normally done in player modelling studies \cite{yannakakis2018artificial,yannakakis2013player}. Second, it introduces a continuous moment-to-moment predictor of engagement in games with a particular application to a popular live streamed game. Finally, the engagement models obtained are highly accurate and general within and across streamers indicating that the function between viewer engagement and gameplay can be learned accurately. Before delving into the details of our methods, the data we solicited, and the key results we obtained, in the next section we elaborate further on our definition of gameplay engagement and review the literature on predictive models of engagement. 

\section{Engagement}

Engagement is a popular yet ambiguous term used in user experience design and research to describe a continuous interest and interaction. The concept of engagement generally encompasses both cognitive and affective processes and is widely associated with attention, arousal \cite{mathur2016engagement}, information interaction \cite{toms2002information}, the flow state \cite{chapman1999engagement}, aesthetics \cite{jennings2000theory}, novelty, and challenge \cite{o2008user}. In the remainder of this section we first review the relationship between viewing behaviour and engagement, we then move onto surveying the links between chat messaging and engagement, and finally we cover core aspects of engagement prediction.

\subsection{Gameplay Engagement via Viewers}

Can viewer behaviour reveal anything about gameplay engagement? According to Yee \cite{yee2006motivations}, the main factors that motivate online gameplay are \emph{immersion, social interest,} and \emph{achievement}. While these factors attempt to describe why people play games, they can also inform us why people watch others playing. Contemporary studies of Sjöblom et al. \cite{sjoblom2017people,sjoblom2017content}, for instance, reveal similar motivations for watching game streams in the form of \emph{affective, social,} and \emph{tension release} needs of viewers.

In this paper we assume there is function between the gameplay state and the engagement of the viewers of that game and we define \emph{engagement} as the active participation of viewers of gameplay. The theoretical grounding of this assumption builds on the \emph{theory of mind} \cite{carruthers1996theories} pointing to our cognitive ability to attribute mental states to ourselves and to others and feel how they might feel. The relationship between a player's and a viewer's engagement has also been described in player experience frameworks such as those of Lazzaro \cite{lazarro2004we}---the \emph{people} factor of ``fun''---or the player archetype taxonomy by Bartle \cite{bartle1996hearts}---the \emph{socialiser} archetype. In practical terms, this attribution of gameplay engagement to viewers can be seen as a form of \emph{third-person} annotation which is the dominant practice for obtaining reliable labels of ground truth in affective computing \cite{calvo2015oxford}.Given the above, the underlying hypothesis explored in this study focuses on people engaging with a game as \emph{viewers} instead of players.

\subsection{From Chat Messages to Engagement}

Although playing is a generally more interactive activity compared to spectating, online viewers are not entirely passive \cite{o2008user}. The participatory communities on Twitch streams, for instance, encourage social engagement while qualitative studies reveal a connection between interaction and pivotal points of streams both in terms of novelty and emotion \cite{hamilton2014streaming}. While spectators react to the stream content often in an emotionally charged manner---producing rapid, unique patterns of crowd communication \cite{ford2017chat,musabirov2018between}---their engagement is entangled with the streamer's focus and the para-social nature of the streamer--spectator relationship \cite{wulf2018watching}. Beyond the suspense of the streamed game content, however, spectators may also engage with the streamer and their online personality.

Intuitively it seems appropriate to associate viewer engagement to high frequencies of chatting behaviour. Recent evidence, however, suggests that higher message frequencies might not always correspond to more engaging content \cite{musabirov2018between}. One explanation of this phenomenon lies within the dynamics between the streamer and the spectators as the continuous interaction of viewers is, in part, mediated by para-social interactions with the streamer \cite{wulf2018watching}. Consequently, when the streamer's attention is directed to the immediate gameplay, spectators lose a point of interaction, causing a drop in their message frequency until a cathartic point is reached, prompting an emotional response. Another explanation lies within the ways viewers manage the incongruity between the novelty of the stimuli and their internal mental models \cite{rauterberg1995framework}.Encountering a boring segment, the viewers' engagement with the video drops; to maintain the level of stimuli, however, the engagement with the chat increases. This can explain the drive behind ``spamming'' behaviour and emoji cascades \cite{barbieri2017towards} which spike during frustrating or boring sections of streams \cite{musabirov2018between}. 

Based on the aforementioned studies we argue that the design of a measurement (or a proxy) of engagement that attributes engagement to higher frequencies of viewer-player interactions can be misleading. In particular, given the dynamics of the examined game---which include long stretches of low tension gameplay (see Section \ref{sec:pubg} for more details)---in this paper we approximate spectator engagement with the gameplay content as a \emph{function inverse to the viewers' chat message frequency}.

\subsection{Continuous Engagement Prediction}

As mentioned earlier, watching streamed gameplay content involves active participation in the form of submitting video recommendations and posting comments. Traditional data analytics methods rely on these metrics---in addition to passive viewership numbers---to calculate the engagement of videos \cite{davidson2010youtube,liikkanen2013three} in terms of dropout, re-engagement, and engagement levels. Dropout and re-engagement can generally be measured and predicted similarly to churn and rely on the user's interaction with a whole platform rather than the streamed content \cite{covington2016deep,wassermann2019machine}. Predicting the engagement level of the video per se, however, often relies on data specific to the streamed content. These metrics focus on the number of interactions during the video such as comments and chat messages \cite{musabirov2018between}. Predictive modelling, in such cases, builds on the language and emotional content of the messages via natural language processing and sentiment analysis \cite{barbieri2017towards}.These approaches may also integrate qualitative analyses via visualisation methods \cite{pan2016twitchviz} or statistical aggregations of chat logs.

Player profiling---a dominant practice in the games industry \cite{el2016game,drachen2018ga}---relies mostly on basic statistical approaches and unsupervised learning methods that derive emergent patterns and distinct groups within the behaviour of players \cite{drachen2009player,bauckhage2015clustering,canossa2018like}.
These methods integrate well into existing industry practices and provide valuable information for both designers and industry stakeholders about how people are interacting with their content in general. Increasingly larger games---with previously unseen amounts of content to stream---come with unique challenges, however, which clustering and profiling methods are unequipped to solve. A response to such growth in available content volume is dynamic player modelling. Player modelling relies on large amounts of data and models the behaviour and experience of players, thereby providing dynamic feedback beyond large-scale observations \cite{yannakakis2013player,yannakakis2018artificial}. In particular, player modelling methods that rely on various supervised learning techniques have already been applied successfully to predict churn \cite{perianez2016churn,viljanen2018playtime}, player behaviour \cite{mahlmann2010predicting,bakkes2012player}, motivation \cite{melhart2019your}, and experience \cite{yannakakis2018ordinal,yannakakis2017ordinal}.

In contrast to the aforementioned studies on media and gameplay engagement, in this paper we focus on a time-continuous prediction of engagement. While traditional analytics focus on evaluating a piece of content (such as a gameplay stream) as a unit we, instead introduce a method for a fine-grained, moment-to-moment prediction of engagement. With this method it is possible to model the moment-to-moment change in engagement---not just highlighting more and less engaging sessions---but providing time-continuous feedback on how engagement changes within a game session. We also introduce a novel proxy for engagement in the form of reverse chat frequency that is used to generate a continuous trace of engagement labels for streamed content. While this ad-hoc metric has to be cross-verified against annotated engagement, our results showcase that features of gameplay content can predict such a measure with supreme levels of accuracy within and across streamers.

\section{Dataset} 

This section outlines the dataset used in this study with an emphasis, on the one hand, on the particular game selected and the telemetry features considered (see Section \ref{sec:pubg}) and, on the other hand, on the engagement annotations we obtained through Twitch (see Section \ref{sec:twitch}).

\subsection{PUBG \& Extracted Features\label{sec:pubg}}

In our attempt to predict engagement via streamed gameplay content we selected PUBG (PUBG Corporation, 2017) as the test-bed for all reported experiments. Our selection is based primarily on two core factors: a) the game's popularity and b) the availability of detailed streaming data. 

\begin{figure}[!tb]
\centering
\includegraphics[width=1.0\linewidth]{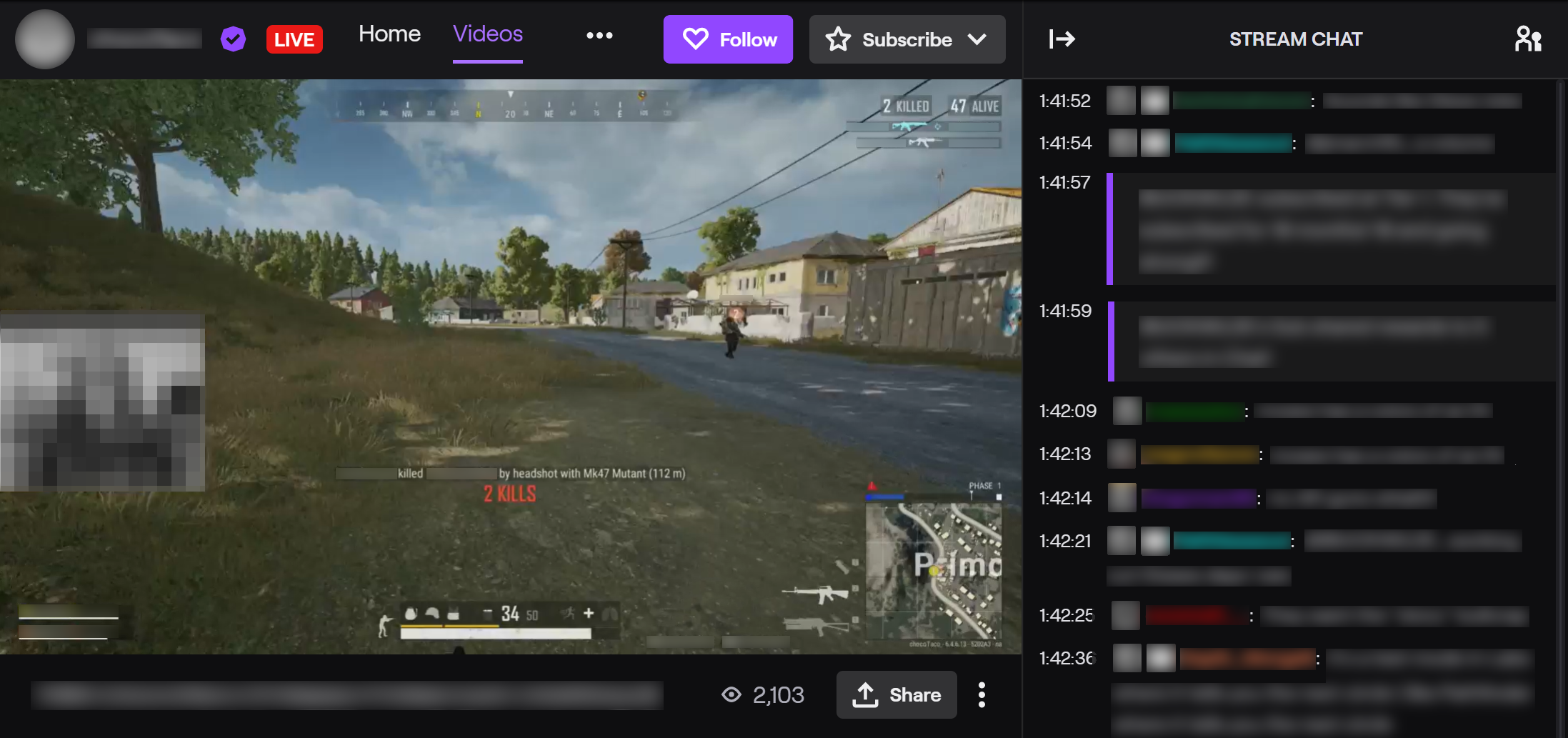}
\caption{To the left: Gameplay stream; to the right: live chat feed. Screenshot of PUBG obtained from Twitch (fair use). PUBG is a registered trademark of PUBG CORPORATION. Identifying information of the streamer and the content of the chat window is blurred out to preserve anonymity.}\label{fig:pubg}
\end{figure}

PUBG (PUBG Corporation, 2017) heralded the rise of Battle-Royale style games and reached high levels of popularity with streamers, who broadcast their gameplay for a wide audience. PUBG is a multiplayer online shooter game, in which a group of players (up to 100 at a time) are dropped into a large open map and left to scavenge for weapons and items, eventually engaging each other in combat until only the winner remains; see Figure \ref{fig:pubg}. The gameplay dynamic is characterised by long stretches of traversal and preparation which are inter-cut by fast bursts of action. As the game progresses, the playable area shrinks, forcing the remaining players closer together, increasing the likelihood of combat. If players remain outside the area of the playable radius they take constant damage; this area is refereed to as the \emph{Blue Zone}. The shrinking of the \emph{Safe Zone} encompassed by the \emph{Blue Zone} is played out in phases. In each phase an \emph{Evacuation Zone} is designated, outside of which players get a \emph{warning} to evacuate the area. The \emph{Blue Zone} then shrinks gradually the \emph{Safe Zone} to the size of the \emph{Evacuation Zone}. The pacing of the game is occasionally broken up by the bombardment of a random localised area, which is indicated by a \emph{Red Zone} and forces players to take shelter inside buildings or evacuate the area. 

PUBG Corporation provides an API and telemetry service\footnote{\url{https://documentation.pubg.com/}}, through which developers and researchers can generate dense datasets of gameplay telemetry. Each session is logged in detail in a hierarchical structure, organised by gameplay events and objects (such as players, pickups, vehicles, and weapons). There are 40 gameplay events and 10 objects available through the API, which cover all players on the level and general game states as well. As this study focuses only on the streamer's content, data relating to other players (e.g., their position, actions, and combat periods which do not involve the streamer) is filtered out. This filtering is also necessary due to the unique structure of \emph{battle royale}-style games. Given the generally large map sizes and the initial scattered distribution of players, one can spend long stretches of the game alone, scavenging for weapons and items, while other players are locked in a battle elsewhere. We only focus on the streamer in an effort to limit the noise of the dataset as most enemy action is completely hidden from the streamer (and their audience) and actual combat happens in short bursts. 

Apart from the aforementioned filtering, we are making use of the full extent of the PUBG API, extracting all event based features without any hand-selection. In particular, we extract 40 PUBG gameplay features for the experiments reported in this paper. The features can be broken down to 5 main categories: \emph{Health, Traversal, Combat, Item Use,} and \emph{General Game State}. The \emph{Health} category includes the streamer's \emph{Health Level} and a number of boolean events: \emph{Healing, Reviving, Receiving Revive, Armor Being Destroyed, Made Groggy, Taking Damage,} and \emph{Being Killed}. The \emph{Traversal} category includes the distance travelled since the last event (\emph{Delta Location}), and the \emph{In Blue Zone, In Red Zone, Swim Start, Swim End, Vault Start, Vehicle Ride, Vehicle Leave} boolean game events. The \emph{Combat} category includes the \emph{Shot Count, Damage Done} scalar values and the following boolean features: \emph{Is Attacking, Weapon Fired, Caused Damage, Destroyed Object, Destroyed Armour, Destroyed Wheel, Destroyed Vehicle, Made Enemy Groggy}. The \emph{Item Use} category keeps track of the \emph{Item Drop, Item Equip, Item Unequip, Item Pickup, Item Pickup From Carepackage, Item Pickup From Lootbox, Item Use, Item Attach, Item Detach} boolean events. Finally, the \emph{General Game State} category includes the \emph{Elapsed Time} (in seconds), \emph{Number of Alive Teams} and \emph{Number of Alive Players} and the \emph{Phase} of the game (i.e., \emph{Blue or Red Zone}).

\subsection{Twitch \& Engagement \label{sec:twitch}}

For the purposes of this study we obtained live PUBG gameplay data from Twitch; currently the largest streaming platform. Although Twitch is a general-purpose live-streaming platform, much of the site's traffic is generated by videogame streaming, both casual and competitive. As eSports and game streaming become more and more popular, the need for selecting more engaging streams, or parts of streams, rises. This is especially true to videogame streaming where fast rising trends can upend previously successful genres and new consumer darlings can skyrocket a company. While Twitch connects streamers with viewers, it also provides a platform for viewers to connect with each other. As it can be seen in Figure \ref{fig:pubg} chatting while watching streamers is a large part of the shared experience. Indeed, contemporary studies on the motivation behind Twitch viewership show that the strongest motivations are social, followed by affective and tension release needs \cite{sjoblom2017content}. While viewers do receive some level of gratification from watching streams and engaging with other viewers, cognitive (i.e., learning) and personal integrative (i.e., recognition by peers) needs are less pronounced \cite{sjoblom2017people} in the users of the platform. 

As mentioned earlier in this paper we measure moment-to-moment engagement as the inverse frequency of chat messages in between two consecutive events of the game. This value is computed as the number of chat messages between two consecutive events and normalised between 0 and 1. To account for the reactionary nature of spectator chat, we look at the number of messages not congruently but following gameplay events (i.e., the number of messages between the observed event and the next event). It is important to note that our metric focuses on the \emph{game content-related} engagement of \emph{the spectators}, and not the player's engagement with the game. Following the study of Makantasis et al. \cite{makantasis2019pixels} in this paper we view the prediction of engagement as a binary classification task, in which the objective is to predict ``high'' or ``low'' engagement labels. In particular, we consider \emph{low} and \emph{high} engaging those events with a message frequency higher and lower, respectively, than a selected threshold, $\alpha$.
While it might seem surprising to associate lower frequencies as moments of viewer's high engagement, by qualitatively inspecting the videos we observed that the chat room tends to be more quiet when fast-pace action is happening on the screen (i.e., viewers are paying more attention to the screen) and chat more when there are calmer slow-pace moments (e.g., as a manifestation of boredom).

\subsection{Streamer Data Collection \& Preprocessing \label{sec:data_analysis}}

\begin{table*}[!tb]
    \caption{Rank, number of videos, number of matches, average number of views (per video), average match duration (in seconds, per match), average number of chat messages (per match), and number of events (per match) across the five selected streamers. Standard deviation is shown in parentheses.}
    \label{tab:basic_stats}

    \centering
    \resizebox{\linewidth}{!}{%
    \begin{tabular}{@{\quad}c@{\quad}|@{\quad}c@{\quad}@{\quad}c@{\quad}@{\quad}c@{\quad}@{\quad}c@{\quad}@{\quad}c@{\quad}@{\quad}c@{\quad}@{\quad}c@{\quad}}
        \hline \hline
        \textbf{Streamer}  & \textbf{Rank} & \textbf{\# Videos} & \textbf{\# Matches} & \textbf{\# Viewers} & \textbf{Duration} & \textbf{\# Chat} & \textbf{\# Events}  \\ \hline \hline
        A         & 1         & 8         & 74        & 3789.6 (195.2)  & 478.2 (516.9) & 279.7 (363.5) & 290.8 (298.6)\\ \hline
        B            & 2         & 2         & 48        & 2150.0  (861.3)   & 636.4 (572.5) & 261.9 (329.8) & 456.2 (392.3) \\ \hline
        C         & 7         & 3         & 89        & 460.6  (35.3)   & 512.7 (558.9) & 91.6  (104.2) & 382.6 (372.8) \\ \hline
        D          & 14        & 3         & 34         & 893.0  (342.5)  & 813.6 (563.4) & 129.1 (139.6) & 387.6 (289.3)\\ \hline
        E            & N/A         & 3         & 79        & 1175.3 (334.4)   & 429.9 (402.6) & 92.8 (110.9) & 369.9 (359.4) \\ \hline
        \hline
        \textbf{Average}  & 6.0 (5.9) & 3.8 (2.4) & 64.8 (22.9) & 1693.8 (1325.8)  & 574.2 (154.1) & 171.0 (92.5) & 377.4 (58.9) \\ \hline \hline
    \end{tabular}}
\end{table*}

To test to which degree we can predict the PUBG engagement through telemetry events, we solicit in-game events and corresponding chat messages from the PUBG API and Twitch API, respectively, from 23 August 2019 to 12 January 2020. In particular we collected data from five anonymised streamers---referred in this paper as \emph{A}, \emph{B}, \emph{C}, \emph{D} and \emph{E}---based on their popularity and the availability of datasets which are large enough to be explored through machine learning.
Table \ref{tab:basic_stats} presents the streamers' ranking \footnote{Ranked by the total viewership hours (hours live $\times$ average viewers) obtained obtained at the time of writing from https://www.twitchmetrics.net/. Only English language speakers are considered in this ranking.}, the number of videos and matches collected, the average number of viewers \footnote{Live value; last accessed at the time of writing.}, the average duration, the number of chat messages, and the number of events collected within the selected timeframe, for each of the five streamers. Based on these statistics we can observe directly that the two top ranked streamers, \emph{A} and \emph{B}, have a substantially higher number of viewers and chat messages per match compared to the other three streamers, who have comparable numbers among them. An interesting exception to this popularity ranking is the average match duration of \emph{D} who seems to be playing roughly two times longer than the other streamers.

After the extraction and preprocessing of the input features (see Section \ref{sec:pubg}) and the transformation of the message frequencies into binary labels (see Section \ref{sec:twitch}), we obtain a total of $119,345$ labelled events. Independently of the class splitting threshold ($\alpha$) value chosen, the dataset presents a highly unbalanced ratio between the two classes, with a majority of the labels being classified as \emph{high} engagement. To balance the dataset, we oversample---by randomly sampling the available samples with replacement---and undersample---by randomly selecting a given number of samples---the minority and majority classes, respectively, resulting to baseline accuracies of 50\%. 
We follow this process individually for the training and test sets so that we eliminate any data leakage. Ideally the over and undersampling method proposed could be isolated on the training set; doing so, however, would yield highly unbalanced test sets that would not ease the analysis in this study. As long as the data processing method we followed does not allow for data leakage between training and test partitions the accuracy of the models obtained is generalisable to potentially highly-unbalanced unseen data.  

\section{Experiments}

For all experiments included in this paper we employ artificial neural networks (ANNs) as our prediction models. We picked ANNs in this initial study because of their evidenced training efficiency in large-scale datasets compared to other machine learning approaches such as support vector machines \cite{abadi2016tensorflow}. More importantly for this work (and potential future studies), ANNs can be easily extended to fuse different input modalities (e.g., pixel and telemetry data), which cannot be easily accomplished with other machine learning techniques \cite{liapis2019modelling}. A number of different ANN architectures and set of hyperparameters have been tested. In particular, we performed a sensitivity analysis across three different parameters: learning rate, number of hidden nodes and dropout rate. Based on that extensive parameter tuning process the ANNs we use feature a single fully-connected hidden layer composed of 128 nodes, followed by a dropout layer \cite{srivastava2014dropout}; the network has an output node that predicts \emph{high} (1) or \emph{low} (0) engagement. All nodes use the ELU activation function \cite{clevert2015fast}, the learning rate equals $10^{-5}$, and the ANN is trained for $100$ epochs. 

In the first round of experiments (Section \ref{sec:single}), we train and test our model individually on each of the five streamers. In the second set of experiments (Section \ref{sec:all_streamers}), we test the scalability of our engagement models across all the streamers. In Section \ref{sec:clustering} we, instead, identify and model the different play styles across all streamers and finally in Section \ref{sec:analysis} we discuss qualitatively about the engagement lines produced from our models across two representative videos.  

\subsection{Individual Streamer Models}\label{sec:single}

In this first set of experiments, we collect and machine learn data coming from each streamer individually. We validate our models using a 5-fold cross-validation scheme in which the matches are distributed randomly within the folds.

To assess which splitting criteria lead to the best model performances, we explore four different threshold $\alpha$ values ($0.0$, $0.1$, $0.2$, $0.3$). Earlier work, however, suggests that this naive approach may lead to split criteria biases, as the model may learn to classify high and low engagement based on trivial differences in the frequency of the events \cite{yannakakis2018ordinal,yannakakis2017ordinal, martinez2014don, makantasis2019pixels}. To address this challenge, we employ an \emph{uncertainty bound} ($\epsilon$) when we split the data so that we filter out any unambiguous datapoints close to the selected threshold value; in particular, we omit all the events that fall within the range $\alpha \pm \epsilon $. In addition to the four $\alpha$ values we explore three different values for $\epsilon = \{0.02, 0.05, 0.08\}$, we examine all the possible combinations of $\alpha$ and $\epsilon$ exhaustively, and we select the configuration with the highest 5-fold cross-validation accuracy. Table \ref{tab:streamer_conf} shows the setup selected for each streamer.

\begin{table}[!tb]
    \caption{Best configurations of $\alpha$ and $\epsilon$ values for each streamer (see Section \ref{sec:single}) and cluster (see Section \ref{sec:clustering}).}
    \label{tab:streamer_conf}
    \centering
    \begin{tabular}{c|ccccc|ccc}
        \hline \hline
        & \multicolumn{5}{c|}{Streamer} & \multicolumn{3}{c}{Cluster} \\
        & A & B & C & D & E & Noob & Explorer & Pro\\ \hline \hline
        $\alpha$    & 0.3 & 0.3 & 0.3 & 0.3 & 0.3    & 0.2 & 0.3 & 0.3\\ 
        $\epsilon$  & 0.0 & 0.05 & 0.05 & 0.02 & 0.08    & 0.00 & 0.05 & 0.08\\ \hline \hline
    \end{tabular}
\end{table}

\begin{figure}[!tb]
\centering
    \includegraphics[width=1.0\linewidth]
    {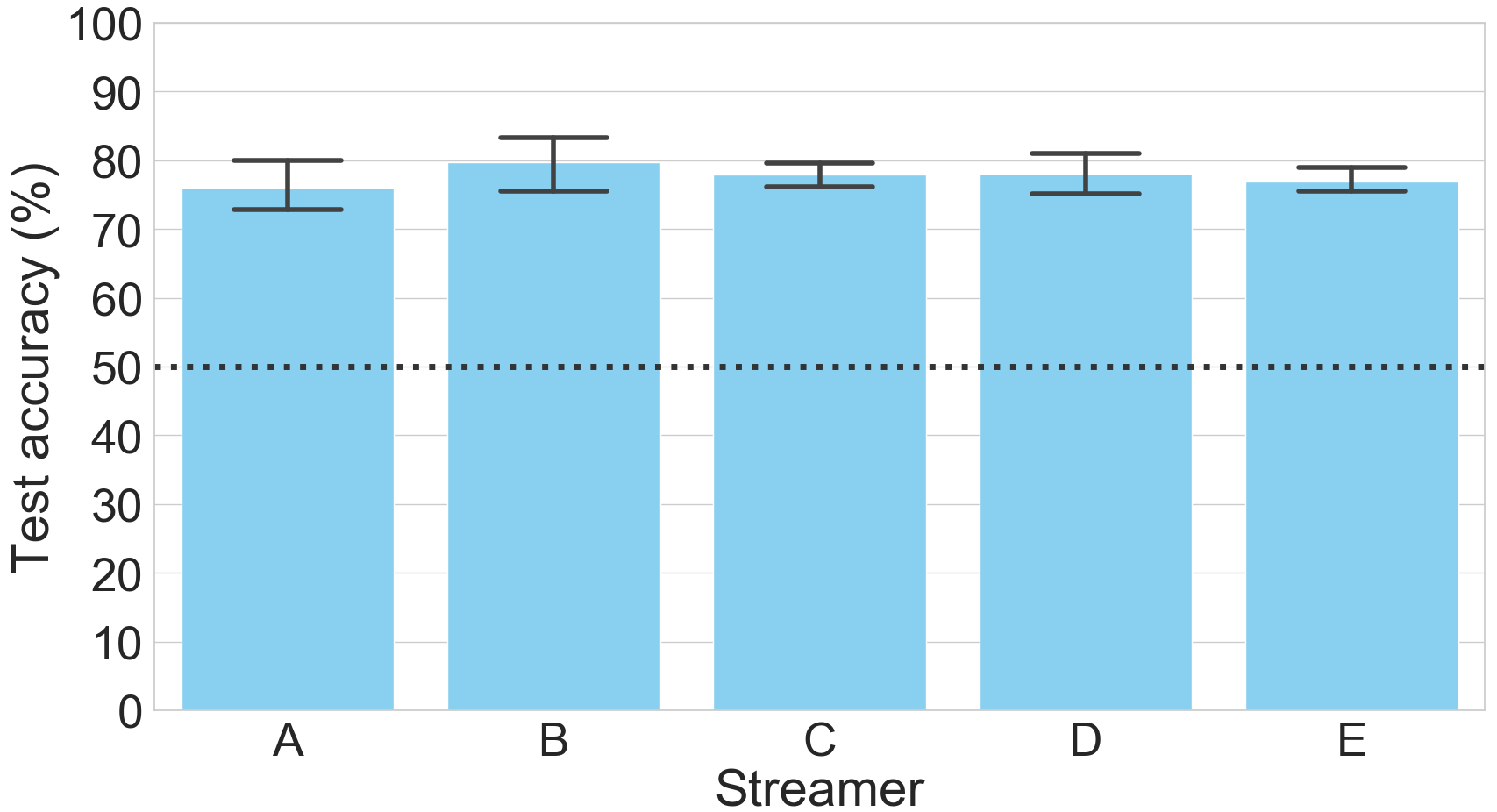}
    \caption{Individual-streamer engagement models: Average 5-fold cross validation accuracies. Error bars denote 95\% confidence intervals.}
\label{fig:accuracies}
\end{figure}

Figure \ref{fig:accuracies} shows the performances achieved for each streamer for the selected hyper-parameters. All individual streamer models of engagement achieve similar performance which reaches 76\% to 80\% on average. In particular the best accuracies are observed for the streamers \emph{B} (79.7\% on average; 84.3\% at best), \emph{D} (78.0\% on average; 82.4\% at best), and \emph{C} (77.8\% on average; 80.43\% at best) while slightly lower values are obtained with \emph{E} (76.8\% on average; 80.8\% at best), and \emph{A} (76.0\% on average; 83.2\% at best). These results already indicate that our method can capture the relationship between streamer telemetry and viewer engagement with a very high accuracy across four different streamers.

\subsection{Models Across All Streamers \label{sec:all_streamers}}

\begin{table}[!tb]
    \caption{Engagement models across all streamers: Average accuracies and their 95\% confidence intervals for different splitting criteria ($\alpha$) and uncertainty bound ($\epsilon$) configurations. The highest accuracy appears in boldface.}
    \label{tab:setup_selection}
    \centering
    \begin{tabular}{c|cccc}
        \hline \hline
        & \multicolumn{4}{c}{$\alpha$}\\

        $\epsilon$  & 0.0 & 0.1 & 0.2  & 0.3\\ \hline \hline
         0.0  & 70.1\% $\pm 1.5$\% & 70.2\% $\pm 3.2$\%  & 72.8\% $\pm 1.4$\% & \textbf{74.7\% $\pm 3.6$\%} \\
         0.02 & 70.2\% $\pm 2.4$\% & 71.1\% $\pm 3.6$\%  & 73.1\% $\pm 1.3$\% & 70.1\% $\pm 11.8$\% \\ 
         0.05 & 70.3\% $\pm 2.0$\% & 71.9\% $\pm 3.4$\%  & 74.1\% $\pm 2.6$\% & 71.2\% $\pm 11.8$\% \\ 
         0.08 & 70.4\% $\pm 1.8$\% & 73.4\% $\pm 3.3$\%  & 74.2\% $\pm 3.2$\% & 71.2\% $\pm 11.9$\% \\ \hline \hline
    \end{tabular}
\end{table}

The findings of the previous set of experiments showcase that capturing the engagement of individual streamers is possible with a very high accuracy. In this section we examine to which degree the models can generalise further and capture the engagement values of unseen streamers.
To test the models' generality we employ the demanding \emph{leave-one-streamer-out} cross-validation scheme \cite{makantasis2019pixels}, in which we train our model based on the data collected from four streamers, and we test it against the remaining streamer. This process is repeated five times, one for each streamer, and the results are averaged. The results obtained with this validation method are a robust indicator of the generalisation capacity of the proposed method, as by subdividing the data into the five available streamers we cannot overfit to a particular streamer data distribution (given the different number of viewers for each streamer) and data leakage is avoided by design.

For all the reported experiments (Table \ref{tab:setup_selection}), we select the best parameter setup based on an exhaustive search of all combinations of $\alpha$ and $\epsilon$ values as performed in Section \ref{sec:single}.  
The best model we could find (74.7\% on average; 78.7\% at best) yields a lower accuracy compared to the accuracies of the models tested on the data of individual streamers. This is unsurprising as a model's generality within-streamer is far easier to achieve than a model's generality across-streamers.

\subsection{Models of Streamer Play Styles}\label{sec:clustering}

Given the results obtained in the first two rounds of experiments it becomes apparent that a general model of engagement across streamers is a rather challenging task. Our hypothesis is that streamers depict varying (non-consistent) behaviours across the matches they play which, in turn, makes any attempt to model engagement across them very challenging for machine learning. We assume, instead, that there are general patterns of play across streamers that machine learning could capture and associate to engagement in an easier manner.

To investigate whether the five streamers show different play styles, we cluster the data collected. The raw data used in the moment-to-moment engagement prediction, however, is too sparse to extract any meaningful clusters. Therefore we first aggregate the $119,345$ events to $324$ matches---i.e., we sum the boolean events (e.g., \emph{Healing}) and we average the scalar values (e.g., \emph{Delta Location})---and for each match we normalise the data with min-max normalisation.

To determine the number of clusters present in the data we follow the approach proposed in \cite{drachen2009player}, we employ two different clustering algorithms---$k$-means and hierarchical clustering \cite{ward1963hierarchical}---and we test the consistency of their outcomes.
We first apply $k$-means to the normalised data for $k$ ranging from 1 to 10, and we compute the quantisation error---i.e., the sum of the distances of every data point to the corresponding cluster centroid. The results show that the percent decrease of the quantisation error when $k$ increases is particularly high with two and three clusters, with a decrease of 53\% and 20\%, respectively. With higher values of $k$ ($k\geq 4$) the difference is more contained (between 1\% and 10\%). Similar results are obtained with the \emph{silhouette coefficient} method \cite{rousseeuw1987silhouettes}. The silhouette coefficient ($s$) is equivalent to the difference of the mean intra-cluster distance and the mean nearest-cluster distance; higher silhouette coefficient values correspond to better defined clusters, bounded between $1$ and $-1$. The results show that for $k = 2$, $k = 3$ and $k = 4$ we obtain the highest coefficients, with $s=0.45$, $s=0.28$ and $s=0.26$ respectively; higher values of $k$, instead, produce lower silhouette coefficients, between $s=0.18$ and $s=0.2$.

\begin{figure}[!tb]
	\centering
		\includegraphics[width=0.9\linewidth]{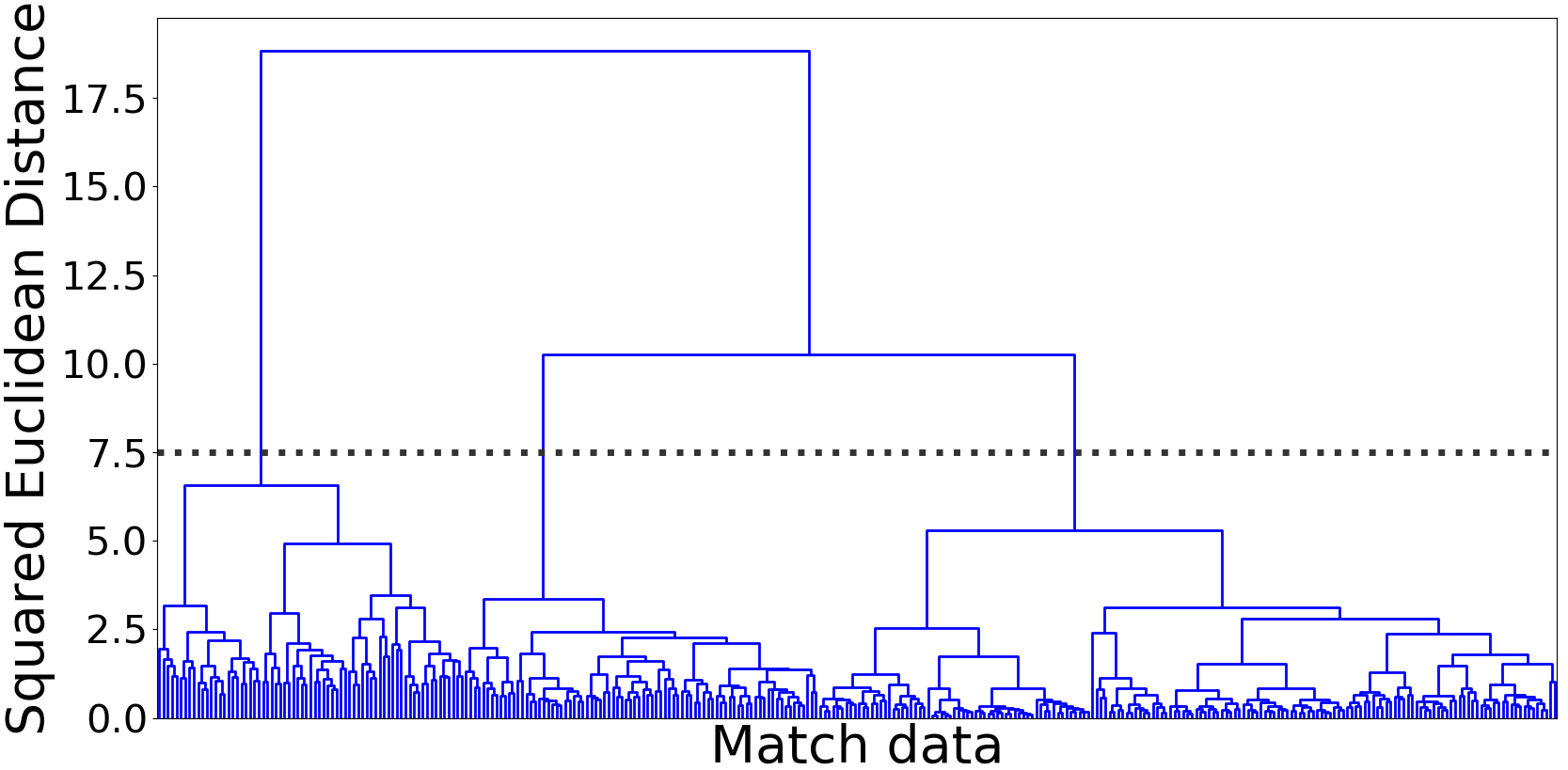}\caption{Dendrogram resulting from hierarchical clustering with the \emph{Ward} method. Indicatively, a distance threshold of 7.5 yields 3 clusters.}
	\label{fig:cluster_analysis}
\end{figure}

An alternative approach to find the appropriate number of clusters is to partition the data in a hierarchical manner starting from every single match and then observe the relationship between the number of clusters and the corresponding squared Euclidean distance that separates those clusters. In our application of hierarchical clustering we use the \emph{Ward} distance metric \cite{ward1963hierarchical}, which minimises the total within-cluster variance. This approach yields comparable results to $k$-means: the dendrogram of Figure \ref{fig:cluster_analysis} shows that a squared Euclidean distance threshold higher than $6.6$ yields three clusters, while a threshold higher than $10.3$ yields two clusters. The analysis performed with these two unsupervised learning algorithms collectively indicates that the most appropriate number of data clusters lies between two and three. Two clusters partition the data into highly unbalanced clusters, with $86$ matches ($74,947$ events) for the first cluster and $238$ matches ($44,398$ events) for the second cluster. Similarly, four clusters yield an unbalanced distribution, with $152$ matches ($14,266$ events) for the first cluster, $53$ matches for the second cluster ($47,088$ events), $95$ for the third cluster ($35,367$ events) and $24$ for the fourth cluster ($22,624$ events).
Three clusters, however, yield a more uniformly distributed match data partitioning, with $155$ ($14,858$ events), $105$ ($42,878$ events) and $64$ matches ($61,609$ events) for the first, second and third cluster, respectively. If we use the information entropy ($H$) \cite{shannon1948mathematical} as a measure of the balance of the distribution of the matches obtained, we notice a higher entropy ($H = 0.95$) with three clusters compared to two ($H = 0.84$) and four clusters ($H = 0.87$). Given the high imbalance of matches partitioned with two clusters, and the similarity of results obtained by the two clustering algorithms it appears that the most reliable way to partition this dataset is through three clusters.

\begin{figure}[!tb]
\centering

  \centering
  \includegraphics[width=1.0\linewidth]{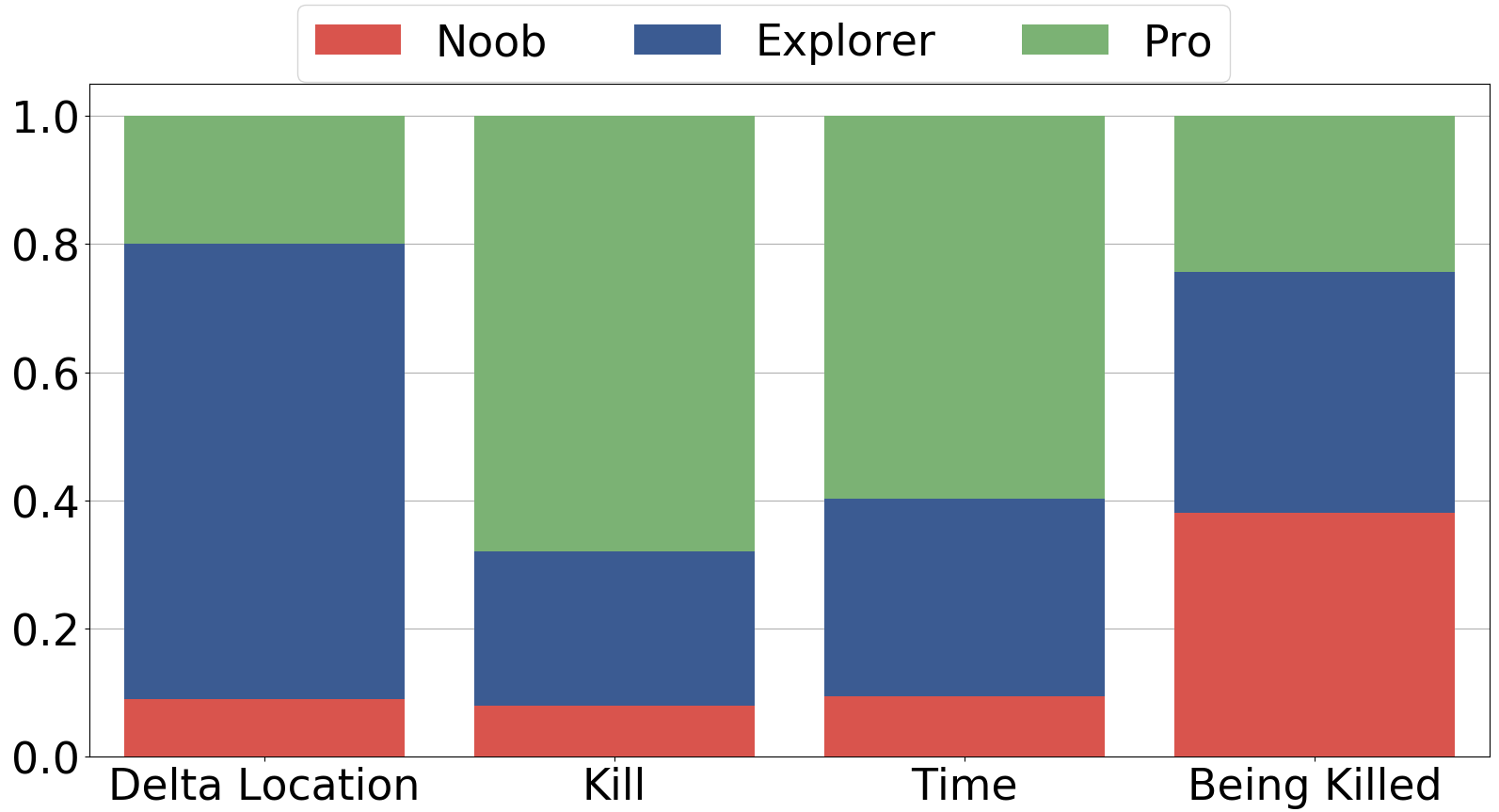}
  \label{fig:deltalocation}
   \caption{Normalised stacked bar chart: mean values of four representative features across the three play styles. \emph{Noob}, \emph{Explorer} and \emph{Pro} are depicted in red, blue and green, respectively.} 
\label{fig:clustered_features}
\end{figure}

To label the three player styles clustered, we investigate how the features of gameplay are grouped within each cluster. Figure \ref{fig:clustered_features} shows the distribution of four representative features across the three clusters. The features displayed are  \emph{Delta Location} (distance covered in a match), \emph{Kill} (number of opponents killed in a match), \emph{Taking Damage} (damage taken by the player in a match), and \emph{Time} (match duration in seconds). Using popular game culture terminology we label the first cluster as \emph{Noob} play style as in those matches the streamer does not play particularly well, he reaches a low number of kills and is more likely to be killed. Meanwhile, the matches are much shorter, most likely because the streamer dies within the first minutes of the match. The second cluster of play style is labelled as \emph{Explorer}: in those matches the streamer explores the map far more---as the \emph{Delta Location} feature is higher compared to the other two clusters---but the performance of the player is still average, as shown by the \emph{Kill} and \emph{Being Killed} features.
Finally, we label the third play style as \emph{Pro} as it features matches where the streamer has played his best: he tends to kill more players, to die less often compared to the other two clusters, and while it takes a considerable amount of damage he survives longer (i.e., higher \emph{Time} values), most likely winning the match.

\begin{figure}[!tb]
\centering
    \includegraphics[width=1.0\linewidth]{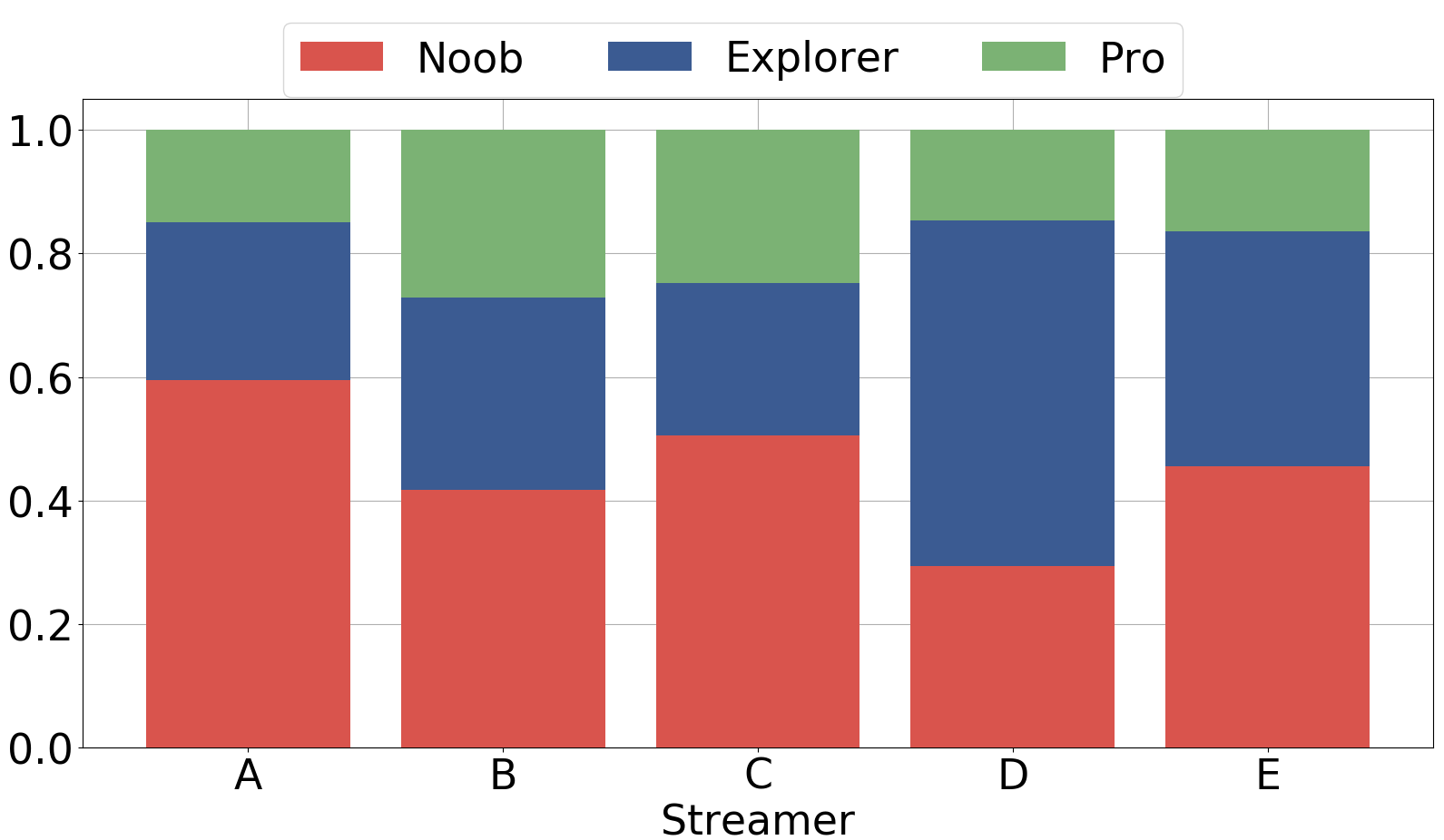}
    \caption{Normalised stacked bar chart of the three play styles across the five streamers.}
\label{fig:player_clustering}
\end{figure}

Figure \ref{fig:player_clustering} shows the distribution of the three play styles across the five streamers and the variation of play styles the different streamers depict, validating our hypothesis. In particular, \emph{A} is labelled as a \emph{Noob} in the majority of his matches, while \emph{D} appears to be more of a \emph{Explorer} player type. \emph{C}, \emph{B}, and \emph{E} show a more uniform distribution of the three play styles in their gameplay. 

\begin{figure}[!tb]
\centering
    \includegraphics[width=1.0\linewidth]{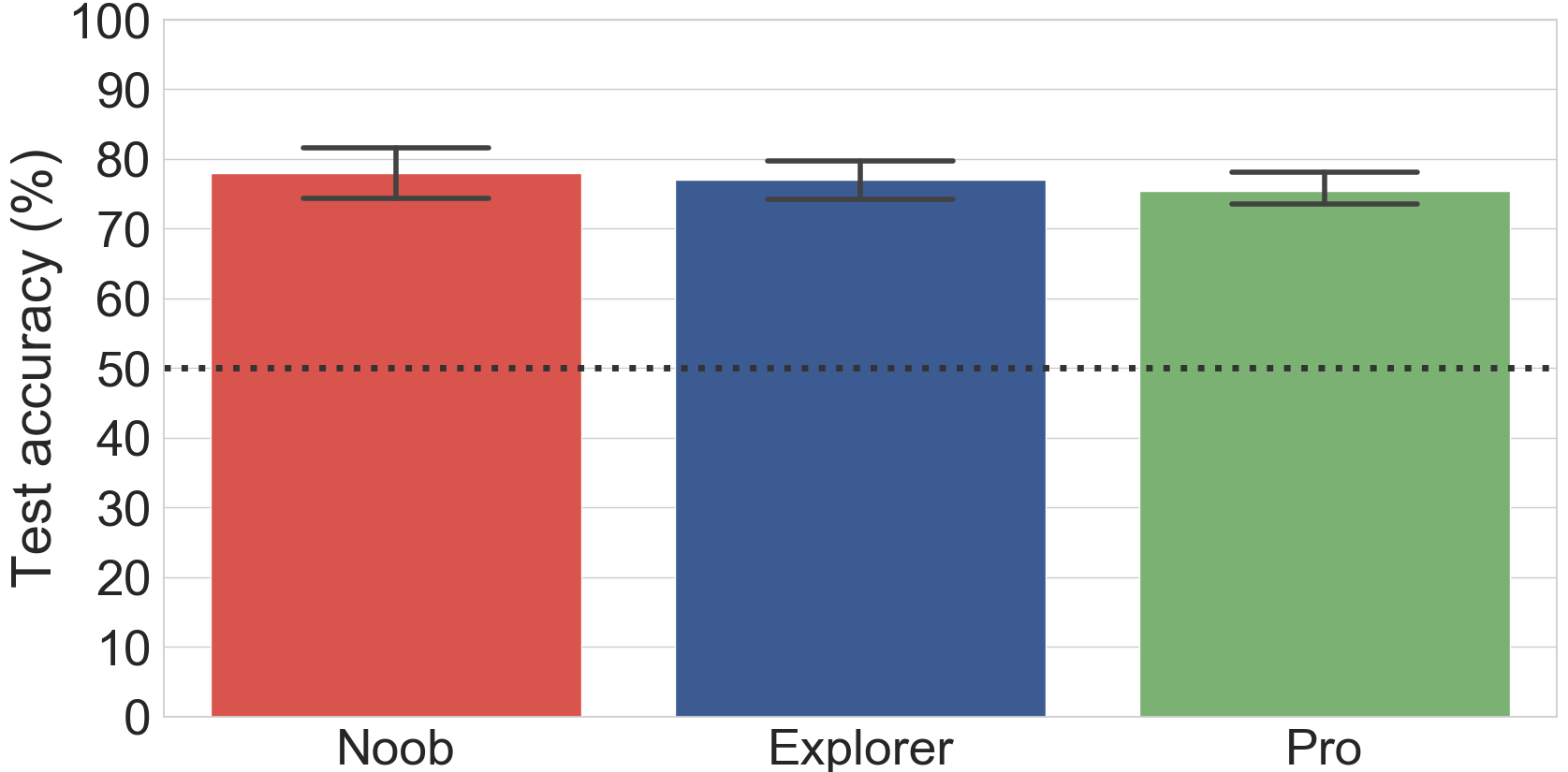}
    \caption{Test accuracies across the three play styles averaged from 5 runs of leave-one-streamer-out validation. Error bars denote 95\% confidence intervals.}
\label{fig:cluster-accuracies}
\end{figure}

Given the three different play styles we obtained we test the generalisability of moment-to-moment engagement models that are built on the play styles, instead of the streamers. We thus train a separate engagement model per play style. Following Section \ref{sec:all_streamers}, we perform an exhaustive search of the predetermined values of $\alpha$ and $\epsilon$ for each play style model. To compare the results obtained, we validate our models using a \emph{leave-one-streamer-out} cross-validation scheme. Figure \ref{fig:cluster-accuracies} illustrates the average test accuracies obtained for the three different play style models of engagement. All models are predicting engagement with high degrees of accuracy (over 75\% on average) but the model for the \emph{Noob} play style performs better (78.0\% on average, 84.2\% at best) than the models for the \emph{Explorer} (77.0\% on average, 81.4\% at best) and the \emph{Pro} play style (75.4\% on average, 80.7\% at best). The key findings in this section suggest that constructing models of engagement across streamer play styles instead across streamers offers a higher generalisability potential for the model.

\subsection{Engagement Line Analysis}\label{sec:analysis}

\begin{figure*}[!tb]
\centering
    \begin{subfloat}
      \centering
      \includegraphics[width=1.0\linewidth]{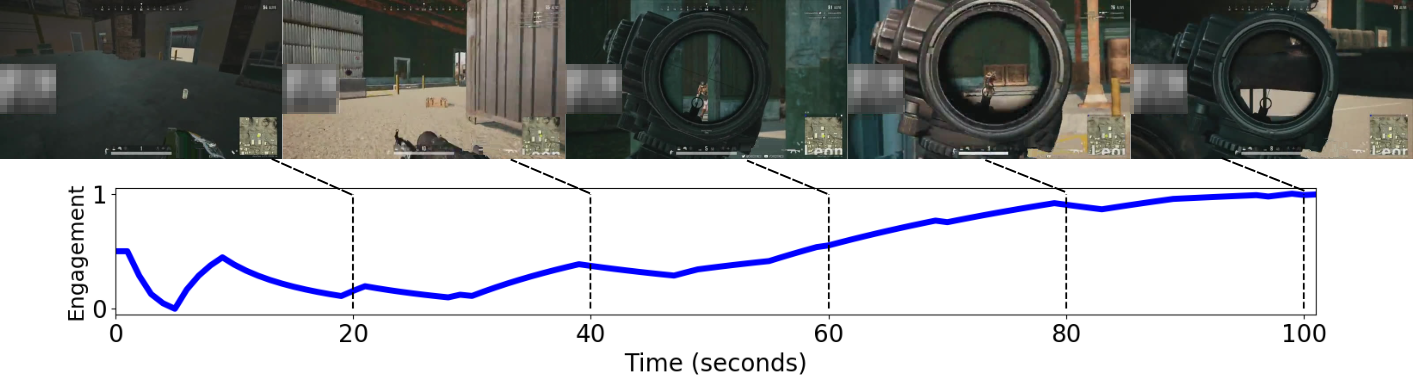}
      \label{fig:video_1}
    \end{subfloat}
    \begin{subfloat}
      \centering
      \includegraphics[width=1.0\linewidth]{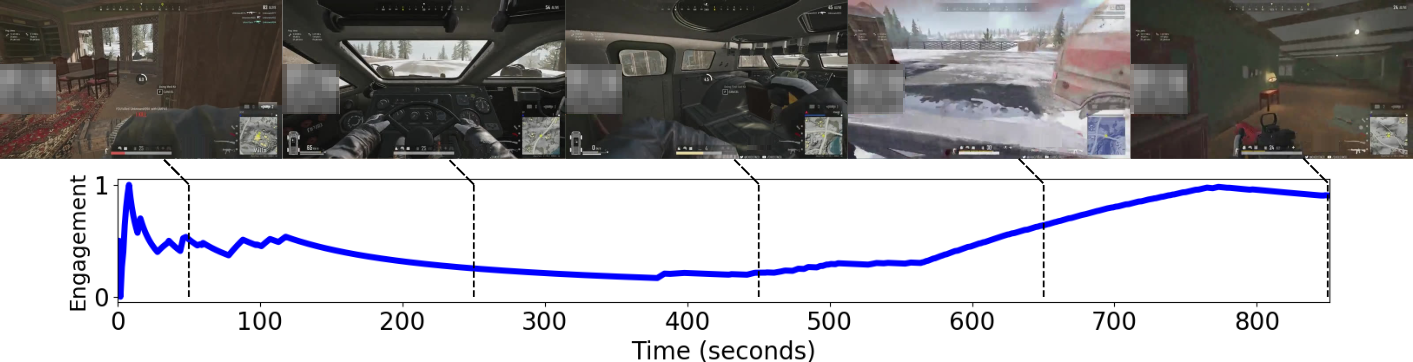}
      \label{fig:video_2}
    \end{subfloat}

    \caption{Two indicative examples of video streams taken from the streamer \emph{A} with the corresponding engagement lines as predicted by our ANN model. The figure depicts stream screenshots at particular events during the game. The video on the screenshots is obtained through the Twitch Developer API (fair use), PUBG is a registered trademark of PUBG CORPORATION.}
\label{fig:timeline}
\end{figure*}

In this section we discuss two indicative examples of PUBG matches with their corresponding engagement line prediction. Figure \ref{fig:timeline} shows the two example engagement lines as associated with video frame sequences taken from streamer \emph{A}. To extract and display a continuous line of engagement between events, we apply a moving average (sampled every second) to the output of the engagement model.

At the top graph of Figure \ref{fig:timeline} we observe a steady increase of engagement as the player starts the match searching for enemies and exploring around the map but with no battle events occurring. Towards the end of the match we observe a further increase of the engagement value which is associated with a fast-pace action phase in which the player engages in battle with several opponents. The player at the bottom graph of Figure \ref{fig:timeline} is shooting and healing himself during the initial phase of the match (first 20 seconds); as a result the model predicts high engagement values for this initial phase. In the middle phase of the match ($200-500$ seconds) the player drives around the map and hence the model yields low levels of engagement. Towards the end of the match, however, the engagement value increases rapidly as the player gets shot and he is engaged in a battle against another player in a house.

\section{Discussion}\label{sec:discussion}

The key findings of this paper suggest that it is not only possible to rely solely on a number of key gameplay events and predict the level of viewer engagement in a continuous fashion but that it can be done with high levels of accuracy.

The approach, however, needs to be tested across a number of varying properties of the dataset considered. Even though the results obtained on 5 PUBG streamers---and several thousand game events---already support the scalability of the method within this game, a larger dataset across more streamers will make our findings even stronger. In addition to the size of the dataset alternative machine learning methods will need to be tested involving deep learning methods as these are more appropriate for larger datasets. It is already highly encouraging, however, that accuracies of over $80\%$ could be reached with relatively simple and shallow ANN architectures. The type of method will also depend on the types of data the model will be trained on. Future studies will consider the use of natural language processing for the chat boxes, facial expression and speech recognition for the streamer, as well as computer vision methods for the pixels of the video stream---as e.g., in \cite{makantasis2019pixels}---in an attempt to reach more accurate models of engagement. While adding more modalities of input to the predictor of engagement might be beneficial to the accuracy of the models it makes the model more dependable on specific input types and, hence, less versatile. The experiments reported in this paper already suggest that simple telemetry features of the game suffice for the construction of engagement models of high accuracy.

Our notion of gameplay engagement is associated with viewer behaviour and, in particular, with the inverse chatting frequency. While such a proxy of engagement is theoretically grounded, it is supported by recent evidence in the literature, and it can be predicted from gameplay telemetry, other ground truths of engagement are planned to be designed and modelled. Any ad-hoc proxy of engagement---as the one proposed here---will need to be empirically cross-verified against annotation data obtained via video annotation tools such as PAGAN \cite{melhart2019pagan,melhart2019paganB}. It is important to note, however, that verifying the ad-hoc engagement proxy we designed in this initial study is beyond the focus of this paper; the core outcome of this study, instead, is that engagement (as defined here) is both theoretically supported and can be predicted accurately from gameplay characteristics in a moment-to-moment fashion. Given the absence of any engagement annotation or emotion labelling in the PUBG game (and most streamed games), reframing the problem of engagement and looking at it from the lens of the viewers' behaviour offers a general-purpose ground truth that can be easily obtained without further human intervention and tedious annotation processes. 

The presented results are relevant to researchers and game industry stakeholders alike. The presented methodology can serve as a basis for future studies towards a more holistic understanding of engagement in games and beyond. Since our ad-hoc metric focuses on game content-related engagement of spectators, it highlights the elements of the gameplay experience, which can be controlled by developers. Thus, games that are developed with streaming content in mind can largely benefit from this type of engagement approximation from the early stages of creation. Our system could also aid industry stakeholders and streaming services who face challenges of information-overload and are in need of algorithmic ways to sort and highlight engaging content. Finally, predicting spectator engagement with the streamed game content can also be used for the procedural generation of play with the aim of creating and curating artificial streams \cite{thawonmas2017ai}. In light of the highly promising results, this study offers some initial evidence that engagement of gameplay videos can be predicted through game telemetry in a particular game of a particular genre. Future studies will focus on testing the proposed methodology across different games of the battle royale genre and also across dissimilar game genres. 

\section{Conclusions}

In this paper we reframe the way gameplay engagement is naturally viewed: from a first person to a third person perspective. In particular, we attempt to predict the moment-to-moment engagement of viewers of live streamed games through their chatting activity. We test our hypothesis that gameplay can be a good predictor of viewer chat frequency in PUBG (PUBG Corporation, 2017) live streams that are obtained over 5 Twitch streamers. We employ shallow ANNs and we model engagement as a function of player metrics across critical events of the game. Our results showcase that modelling engagement in a continuous fashion is not only possible but that viewer engagement can be predicted with accuracies, that reach 80\% on average (and 84\% at best). The models appear to be versatile within and across different streamers as well as across different play styles of streamers. This initial study showcases the potential of the approach for measuring moment-to-moment engagement in game streams (and beyond) through simple yet critical events in the game. 

\bibliographystyle{ACM-Reference-Format}
\bibliography{pubg}

\end{document}